# Representation of Uncertainty in Electric Energy Market Models: Pricing Implication and Formulation

Mohammad Ghaljehei, *Student Member, IEEE,* Mojdeh Khorsand, *Member, IEEE*

*Abstract*—Modern market management systems continue to evolve due to the intentions to improve system security and reliability. This evolvement has been leading to a transition of market auction models from a deterministic structure with approximations on the reliability criteria (e.g., acquirement of contingency reserve through proxy reserve policies) to explicit representation of contingencies (e.g., estimation of post-contingency states via participation factors and stochastic programming). This paper proposes a comprehensive framework to establish various procedures for evaluating: (i) transparency and incentive compatibility of different contingency modeling approaches, and (ii) efficiency of two possible stochastic market designs. First, the concept of *securitized LMP* is presented to solve the issue of how market participants should be compensated for providing *N*-1 reliability services. Then, pricing implications and settlements of three market models are compared: (i) a deterministic market model with proxy serve policies, (ii) state-of-the-art market models with estimated post-contingency states, and (iii) a two-stage stochastic market model. Second, this paper evaluates two stochastic market models while accounting for potential adjustments from day-ahead scheduling to real-time operation: (i) minimizing expected operating cost of all *N*-1 scenarios, and (ii) minimizing the base-case (or no contingency) cost. These analyses are conducted on IEEE 118-bus test system.

*Index Terms*— Locational marginal price, market design, market settlements, *N*-1 reliability, stochastic market model.

## Nomenclature

*Sets and Indices*
| | |
|---|---|
| $c$ | Index of operating state; 0 for the base-case, non-zero for contingencies. |
| $C_0, C_g, C_k$ | Set of scenarios representing base-case, generator, and line contingencies, respectively. |
| $g$ | Index of generators, $g \in G$. |
| $g(n)$ | Set of generators connected to node $n$. |
| $k, \ell$ | Index of transmission lines, $k, \ell \in K$. |
| $n$ | Index of buses, $n \in N$. |
| $t$ | Index of time periods, $t \in T$. |

*Parameters*
| | |
|---|---|
| $c_g^{NL}, c_g^{SD}, c_g^{SU}$ | No-load, shutdown and startup costs of unit $g$. |
| $Load_{nt}$ | Load at bus $n$ at time period $t$. |
| $\bar{P}_g, \bar{u}_g, \bar{r}_g$ | Day-ahead scheduled power output, commitment, and contingency reserve of unit $g$. |
| $P_g^{max}, P_g^{min}$ | Maximum output and minimum output of unit $g$. |
| $PTDF_{cnk}^{ref}$ | Power transfer distribution factor during operating state $c$ for line $k$ for an injection at $n$. |
| $LODF_{k\ell}^{ref}$ | Line outage distribution factor representing the change in flow on line $k$ for outage of line $l$. |
| $c_g^p$ | Variable cost of unit $g$ (\$/MWh). |
| $N1_k$ | $N$-1 contingency indicator of transmission line $k$; 0 for a contingency on line $k$; otherwise, 1. |
| $N1_g$ | $N$-1 contingency indicator of generator $g$; 0 for a contingency on generator $g$; otherwise, 1. |
| $R_g^{HR}, R_g^{10}$ | Hourly and 10-min ramp rates of unit $g$. |
| $R_g^{SU}, R_g^{SD}$ | Startup and shutdown ramp rates of unit $g$. |
| $UT_g, DT_g$ | Minimum up time and down time of unit $g$. |
| $\pi_{BC}$ | Probability of base-case operating state. |
| $\pi_c$ | Probability of contingency operating state $c$. |
| $P_k^{max}$ | Thermal rating of transmission line $k$. |
| $P_k^{max,c}$ | Emergency thermal rating of transmission line $k$. |

*Variables*
| | |
|---|---|
| $P_{gct}$ | Output of unit $g$ for operating state $c$ at period $t$. |
| $P_{nct}^{inj}$ | Net power injection at bus $n$ for operating state $c$ at period $t$. |
| $FL^0{}_{lt}$ | Flow on transmission line $l$ at period $t$. |
| $r_{gt}$ | Contingency reserve of unit $g$ at period $t$. |
| $u_{gt}$ | Unit commitment variable for unit $g$ at period $t$. |
| $v_{gt}, w_{gt}$ | Startup and shutdown variables for unit $g$ at period $t$. |
| $d_{nt}$ | Demand at bus $n$ at period $t$. |
| $\lambda_{nct}$ | Locational marginal price at bus $n$ for operating state time $c$ at period $t$. |

## I. Introduction

**E**LECTRIC systems are considered the greatest achievement of the 20th century by the National Academy of Engineering [1]. Operational scheduling of this sophisticated engineering system necessitates consideration of both economical and reliability aspects. However, due to its complexity, it is non-trivial to model all system components, capture detailed characteristics of all system assets, and satisfy all reliability requirements all together. Hence, existing operational scheduling models are designed with approximations, e.g., DC approximation of power flow and approximations of the *N*-1 reliability mandate (i.e., loss of a single element, e.g. a generator or a non-radial transmission asset, should not cause involuntary load shedding [2]). This mandate makes underlying market model stochastic in nature.

Some of the existing electricity market operators solve day-ahead (DA) security-constrained unit commitment (SCUC) models with an approximation of the *N*-1 reliability mandate via a proxy reserve requirement [3], where total of contingency reserves across the power system is forced to be greater than a certain threshold. Such SCUC models do not account for and guarantee post-contingency reserve deliverability.

Furthermore, some other operators, e.g. Midcontinent independent system operator (MISO), model zonal reserve requirements [4]. This model is unable to differentiate the generators within each zone regardless of their ability or inability to deploy reserve due to transmission system congestion. To compensate for the approximations in market models, the operator may intervene and make adjustments in the market solutions. Such interventions are referred to as out-of-market corrections (OMC) [5] or exceptional dispatches [6] which include committing additional generation units or redispatching committed units. After an $N$-1 reliable dispatch solution is obtained, the settlements are calculated. The existing practice to calculate market settlements is to use the locational marginal prices (LMPs) from the DA SCUC (which have not been affected by the OMC) and the modified $N$-1 reliable dispatch solution [6]. Such market models are unable to account for the true value of reserve provided by each generator in a nodal basis and consequently might not incentivize resources to do as directed by the market. These inabilities to impact market prices will lead to a *missing money* problem (i.e., insufficient compensation received by generators) and cause a natural unfairness as market participants might not be dispatched fairly with these pricing schemes.

In the electricity markets, compensation mechanisms are still a subject of debate. Some ISOs have capacity market auctions with estimated peak load and peak period prices in an attempt to compensate market participants for providing reliability services during peak periods [7]. Others, such as MISO, use convex hull pricing, which is an alternative pricing scheme in non-convex markets to clear the market while also minimizing the total uplift payments [8]. While these approaches are developed in an attempt to improve compensation mechanisms, the issue still persists; existing pricing schemes do not sufficiently reflect the true value of providing energy during contingencies, as these uncertain events are not explicitly included in the market models.

Explicit representation of the contingencies via a two-stage stochastic extensive-form SCUC (ESCUC) enables inclusion of value of reliability services into the LMPs and can reduce the *missing money* problem. ESCUC optimizes the recourse decision variables (or corrective actions) while explicitly considering the network constraints for the post-contingency state, which ensures nodal reserve deployment considering physical network limitations. Pricing analyses for stochastic security-constrained approaches in the energy and reserve markets are presented in [9]–[12]. Reference [9] investigates a method to compensate generators for energy and reserve. The authors in [9] derive a pricing mechanism where the generators are compensated for the modeled $N$-1 scenarios. However, results are shown for only a single time period while the formulated model allows for load shedding through a fixed cost. It is worth noting that the fixed cost of load shedding is hard to estimate since (1) it is not necessarily proportional to bids submitted to the market as energy bids, and (2) it is not the same (fixed cost) for different sectors (industry, domestic, commercial). The authors in [10] and [11] formulate a multi-period stochastic SCUC model that takes into account the post-contingency states for pre-selected contingencies, while allowing load-shedding. In [12], the authors utilize a two-stage stochastic linear program to propose different methods to compensate generators. The models presented in [9]–[12] allow for load-shedding through the value-of-lost-load (VOLL); however, this approach is subjective since the obtained results are sensitive to the choice of VOLL. Additionally, a comprehensive economic evaluation (e.g., generation revenue, generation rent, load payment, and congestion rent) for the stochastic two-stage SCUC and its comparison with other contingency modeling approaches have neither been included nor analyzed in prior work.

In addition, prior work proposed approaches based on the estimated post-contingency states using *pre-determined* participation factors. These approaches fill the gap between the traditional deterministic and the stochastic models by explicitly representing contingencies without any second-stage recourse decisions. For instance, line outage distribution factors (LODFs) can be used to explicitly model the transmission line contingencies [13]. Another example is CAISO, which intends to explicitly enforce the post-contingency transmission constraints for the generator contingencies using generator loss distribution factors (GDF) [14]. Also [15] and [16] have proposed a set of $G$-1 security constraints, thereby, contingency reserves are allocated more efficiently in the system with respect to post-contingency dispatch feasibility. With the explicit modeling of contingency events within the state-of-the-art market auction models, the industry is actually moving from the deterministic market models to a stochastic model. With such stochastic modeling, it is desirable for LMPs to reflect the value and quality of services provided by market participants in response to contingencies. However, there are unsolved issues regardless of the choice of uncertainty modeling: *generators compensation for providing N-1 reliability services as well as impact of contingency modeling on prices.*

Apart from the above issue, in the context of stochastic market designs, majority of prior work adopt an objective function that optimizes the base-case along with the expected cost of the post-contingency states [10], [12], [17]–[19]. However, there is a number of reasons why optimizing over an expected cost may not be the best choice. Firstly, during emergency conditions in real-time (RT), the operator may not exactly follow the proposed corrective actions since the intention during an emergency condition is not to minimize cost; rather, the goal is to recover from the event as quickly as possible to prevent future unforeseen problems that could lead to cascading outages. Furthermore, it is difficult to accurately predict the probability of outages, which itself can lead to different pricing implications and market solutions. There are other studies [9], [20], and [21] that minimize the base-case costs as their objective functions, while the model is still a stochastic two-stage SCUC with explicit representation of post-contingency states. References [9], [10], [12], [17]–[21] aim to improve stochastic market models, whereas the proper design of objective function for these models has neither been analyzed nor included in prior work.

The above literature survey reveals a few gaps that need additional attention and further work. To the best of authors'

knowledge, very limited efforts have been done on how various choices of modeling contingency events affect the potential operational efficiency, incentive compatibility, market transparency, and market settlement policies in the markets with inherent stochastic nature. In addition, no prior work has been conducted about how to formulate the objective function that maintains efficiency for the stochastic markets with the uncertain contingency events. The primary contributions of this work are as follows:

- Impacts of contingency modeling strategies on electricity market outcomes, pricing, and settlements are analyzed. This paper leverages the duality theory to calculate LMP in a stochastic market model; this theoretical method confirms that the value of providing *N-1* reliability services can be reflected in the LMPs of such stochastic market models. This pricing scheme is then compared to two state-of-the-art market auction models, where achieving *N-1* reliable dispatch is postponed to OMC. Also, the market settlements of these models are calculated and compared. With these analyses, this paper seeks to inform market stakeholders about the impacts of contingency modeling approaches in the DA market process and their implications on pricing and settlements.
- The choices of objective function for the stochastic market models are analyzed; a stochastic market design with an expected cost objective function is examined and compared with the base-case costs minimization objective function from two aspects: (i) realized cost during *N-1* contingencies and (ii) effects of inaccurate calculation of the probabilities on market outcomes.

It is worth noting that the aim of this paper is not to develop a new market design; it is rather to propose a framework to evaluate existing market models and stochastic market models in terms of potential operational efficiency, incentive compatibility, fair pricing, and transparency.

The rest of the paper is organized as follows. Section II presents model formulation. Section III focuses on pricing implication of contingency modeling approaches. Section IV evaluates choices of objective function for stochastic market frameworks. Finally, section V concludes the paper.

## II. MODEL FORMULATION

Previous studies in the area of managing discrete uncertain events (i.e., contingencies) in the SCUC problem can be categorized as follows: (i) proxy reserve policies, (ii) modeling system response via participation factors, e.g. LODF and GDF, (iii) stochastic programming approaches, e.g., ESCUC, and (iv) chance-constrained optimization and robust optimization. The main focus of this work is on managing uncertainty through (i), (ii), and (iii) above. In the following three subsections, model formulations related to these approaches are presented.

### A. SCUC with deterministic proxy reserve requirement

A SCUC market model with deterministic proxy reserve requirement is presented in (1)-(19), which is similar to the model in [15]. The objective function, minimizing total operating costs, is presented in (1). In this formulation, constraints (2) and (3) model the relationship of the unit commitment variables with the startup and shutdown variables, respectively. Constraints (4)-(7) model the binary commitment ($u_{gt}$) decision and the startup ($v_{gt}$) and shutdown ($w_{gt}$) decisions, respectively. Minimum up and down time constraints are enforced by (8) and (9). Constraints (10) and (11) ensure ramp rate limits. Constraint (12) guarantees balance between the power injection and withdrawal at every bus and constraint (13) ensures the energy balance between load and generation across the system. Constraint (14) models the transmission line limits. The generator output limits are presented by (15) and (16), while constraint (17) limits the spinning reserve to the 10-minute generators' ramp rate capability. Proxy reserve requirements are modeled through (18)-(19).

$minimize \sum_g \sum_t (c_g^p P_{g0t} + c_g^{NL} u_{gt} + c_g^{SU} v_{gt} + c_g^{SD} w_{gt})$ (1)

Subject to

$v_{gt} \geq u_{gt} - u_{gt-1}, \forall g, t \geq 2$ (2)
$w_{gt} \geq u_{gt-1} - u_{gt}, \forall g, t \geq 2$ (3)
$v_{gt} \geq u_{gt}, w_{gt} = 0, \forall g, t = 1$ (4)
$0 \leq v_{gt} \leq 1, \forall g, t$ (5)
$0 \leq w_{gt} \leq 1, \forall g, t$ (6)
$u_{gt} \in \{0,1\}, \forall g, t$ (7)
$\sum_{s=t-UT_g+1}^{t} v_{gs} \leq u_{gt}, \forall g, t \geq UT_g$ (8)
$\sum_{s=t-DT_g+1}^{t} w_{gs} \leq 1 - u_{gt}, \forall g, t \geq DT_g$ (9)
$P_{g0t} - P_{g0t-1} \leq R_g^{HR} u_{gt-1} + R_g^{SU} v_{gt}, \forall g, t$ (10)
$P_{g0t-1} - P_{g0t} \leq R_g^{HR} u_{gt} + R_g^{SD} w_{gt}, \forall g, t$ (11)
$\sum_{g \in g(n)} P_{g0t} - Load_{nt} = P_{n0t}^{inj}, \forall n, t$ (12)
$\sum_n P_{n0t}^{inj} = 0, \forall c, t$ (13)
$-P_k^{max} \leq \sum_n P_{n0t}^{inj} PTDF_{0nk}^{ref} \leq P_k^{max}, \forall k, t$ (14)
$P_{g0t} + r_{gt} \leq P_g^{max} u_{gt}, \forall g, t$ (15)
$P_g^{min} u_{gt} \leq P_{g0t}, \forall g, t$ (16)
$0 \leq r_{gt} \leq R_g^{10} u_{gt}, \forall g, t$ (17)
$\sum_{j \in G} r_{jt} \geq P_{g0t} + r_{gt}, \forall g, t$ (18)
$\sum_g r_{gt} \geq \eta\% \sum_n Load_{nt}, \forall t$ (19)

### B. SCUC with line contingency modeling using LODF

Today, some ISOs use LODF to explicitly model non-radial line contingencies in the DA SCUC model without adding any second-stage recourse variables [13]. The LODFs are participation factors, which indicate redistribution of flow on the transmission lines (e.g., line $k$) after outage of a line (e.g., line $\ell$) [5]. The SCUC model that incorporates explicit representation of the transmission contingency using LODF is presented in (20)-(23).

$minimize \sum_g \sum_t (c_g^p P_{g0t} + c_g^{NL} u_{gt} + c_g^{SU} v_{gt} + c_g^{SD} w_{gt})$ (20)

Subject to

Constraints (2)-(19) (21)
$-P_k^{max,c} \leq \sum_n P_{n0t}^{inj} PTDF_{0nk}^{ref} + LODF_{k\ell}^{ref} FL^0{}_{\ell t} \leq P_k^{max,c},$
$\forall k \neq \ell, t$ (22)
$FL^0{}_{\ell t} = \sum_n P_{n0t}^{inj} PTDF_{0n\ell}^{ref}, \forall \ell, t$ (23)

### C. ESCUC market model

The ESCUC problem is formulated as a two-stage stochastic program. The scenarios represent base-case pre-contingency scenario and contingency scenarios (i.e., the loss of non-radial transmission line and generator) with their corresponding probabilities. This market model is defined by (24)-(35).

$$\text{minimize} \sum_g \sum_t \pi_{BC} c_g^p P_{g0t} + \sum_g \sum_t (c_g^{NL} u_{gt} + c_g^{SU} v_{gt} + c_g^{SD} w_{gt}) + \sum_g \sum_{c \neq C_0} \sum_t \pi_c c_g^p P_{gct} \quad (24)$$

Subject to

Constraints (2)-(11) and (15)  (25)

$$\sum_{g \in g(n)} P_{g0t} - d_{nt} = P_{n0t}^{inj}, \forall n, t \quad [\lambda_{n0t}] \quad (26a)$$

$$\sum_{g \in g(n)} P_{gct} - d_{nt} = P_{nct}^{inj}, \forall n, c \in C_g, t \quad [\lambda_{nct}] \quad (26b)$$

$$\sum_{g \in g(n)} P_{gct} - d_{nt} = P_{nct}^{inj}, \forall n, c \in C_k, t \quad [\lambda_{nct}] \quad (26c)$$

$$d_{nt} = Load_{nt}, \forall g, t \quad [\lambda_{nt}^{securitized}] \quad (27)$$

$$\sum_n P_{nct}^{inj} = 0, \forall c, t \quad (28)$$

$$-P_k^{max} \leq \sum_n P_{nct}^{inj} PTDF_{0nk}^{ref} \leq P_k^{max}, \forall k, c \neq C_k, t \quad (29)$$

$$-P_k^{max,c} N1_k \leq \sum_n P_{nct}^{inj} PTDF_{cnk}^{ref} \leq P_k^{max,c} N1_k, \forall k, c \in C_k, t \quad (30)$$

$$P_g^{min} u_{gt} N1_g \leq P_{gct} \leq P_g^{max} u_{gt} N1_g, \forall g, c, t \quad (31)$$

$$P_{gct} - P_{g0t} \leq R_g^{10} u_{gt}, \forall g: g \neq c, c, t \quad (32)$$

$$P_{g0t} - P_{gct} \leq R_g^{10} u_{gt}, \forall g: g \neq c, c, t \quad (33)$$

$$P_{gct} - P_{g0t} \leq r_{gt}, \forall g: g \neq c, c, t \quad (34)$$

$$P_{g0t} - P_{gct} \leq r_{gt}, \forall g: g \neq c, c, t \quad (35)$$

In the above formulation, the objective is to minimize the expected operating cost over a set of uncertain scenarios as presented in (24). The node balance constraint (see (26a-c)) is separated to distinguish when the constraint represents the base-case (26a), $G$-1 generation contingency scenarios (26b), and finally $T$-1 transmission contingency scenarios (26c). Constraint (28) enforces energy balance between the supply and the demand at the system level. Transmission line capacity limit for the base-case scenario and the $G$-1 generation contingency scenarios is constrained by (29), whereas that for $T$-1 transmission contingency scenarios is imposed by (30). The generator output limit constraint is represented by (31). Finally, deviation of an online generator output level (see (32)-(35)) from the base-case dispatch to the post-contingency dispatch is limited by its reserve ($r_{gt}$) and by its 10-minute ramp rate ($R_g^{10}$). Note that for each scenario, only one of $N1_k$ and $N1_g$ is set to 0 while the rest is set to 1.

### III. PRICING IMPLICATIONS OF CONTINGENCY MODELING APPROACHES

The market model presented in Sections II.A and II.B give market solutions which may not be $N$-1 and $G$-1 reliable, respectively. To achieve $N$-1 reliable solutions, the market operators implement OMC on their market solutions [6]. To replicate this practice, this paper implements OMC on the output of market models from Sections II.A and II. B. The OMC approach used in this work is similar to [6], [15] and [22]. In this approach, the generation units that are committed in the DA SCUC market model are not allowed to be de-committed, and their dispatches are limited to the original approximated DA solution by their 10-minute ramp rate limit. However, modifying the dispatch and the commitment of additional units is allowed in order to ensure reliable operation.

After an $N$-1 reliable dispatch solution is obtained through OMC approach, the settlements are calculated using DA market LMP [6]. However, these settlements may not reflect the true value of $N$-1 reliability services due to the discrepancy between LMP calculation and final dispatch solution. Thus, such practice may not be incentive compatible for market participants, especially those who provide reliability services.

On the other hand, since all contingencies are represented endogenously in the ESCUC market model (Section II.C), the obtained solution is expected to be $N$-1 reliable. For this model, a pricing mechanism can be obtained to properly incentivize all market participants for providing energy and contingency reserve. In this paper, the concept of s*ecuritized LMP* (SLMP) is presented to better capture the value of reliability services. The SLMP is the dual variable of (27), i.e., $\lambda_{nt}^{securitized}$. Since ESCUC model is a mixed-integer linear program, its dual formulation is not well-defined. However, after fixing the binary variables to their values at the best solution found, the linear model of ESCUC is achieved, which has a well-defined dual formulation. Equation (36) is obtained by deriving the dual formulation from the ESCUC linear primal problem, which shows the relationship between LMPs from the base-case (26a), contingency scenarios (26b)-(26c), and the SLMP.

$$\lambda_{nt}^{securitized} = \lambda_{n0t} + \sum_{c \in C_g} \lambda_{nct} + \sum_{c \in C_k} \lambda_{nct}, \forall n, t \quad (36)$$

It is worth mentioning that to obtain the relation presented in (36), the demand is treated as a variable in (26a-c), and the model enforces $d_{nt} = Load_{nt}$ in (27). The first term in the right-hand side of (36) represents energy and congestion components of SLMP in pre-contingency state, while the second and third terms represent energy and congestion components of SLMP in post-contingency states for the generators and non-radial transmission lines contingencies, respectively. Therefore, the SLMP inherently captures the true value of reserves in the post-contingency state on a nodal basis. The pricing scheme presented here from the ESCUC market model has the advantage that it permits the ISOs to gauge how the market participants should be compensated for providing contingency-based reserve.

The market settlements are compared for three market models, i.e., models in Sections II. A, II. B, and II.C. Fig. 1 illustrates the procedure for comparing these market models.

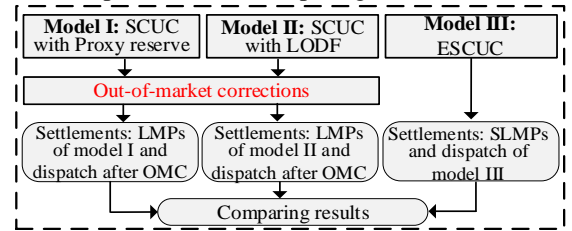

Fig. 1: Procedure for pricing implications comparison of market models.

#### A. Testing & Results of pricing implication

CPLEX v12.8 are used to perform all simulations on a computer with an Intel Core i7 CPU @ 2.20 GHz, 16 GB RAM, and 64-bit operating system. A modified 118-bus IEEE test system [23] is used to implement the market auction models, which has 54 generators, 186 lines (177 non-radial), and 91 loads. Set $C_g$ and $C_k$ include $N$-1 contingencies for all generators and non-radial transmission line elements, respectively. Consequently, there are 232 scenarios modeled in the ESCUC market auction model, including the base-case scenario, 54 generator contingencies and 177 non-radial transmission line contingencies. The probability of contingencies is calculated from historical failure rates [24]. The probability of base-case is considered to be 0.946 (i.e., $\pi_{BC}$=0.946) in order to make the summation of probabilities over all scenarios equals to 1. The relative MIP gap is set to 0%. The three market auction models, i.e., SCUC with proxy reserve requirements (abbreviated as "SCUC-Prxy"), SCUC with transmission contingency modeled using LODF (abbreviated as "SCUC-LODF"), and ESCUC (abbreviated as "SCUC-Extsv"), are compared in terms of operational cost, incentive compatibility, and market settlements.

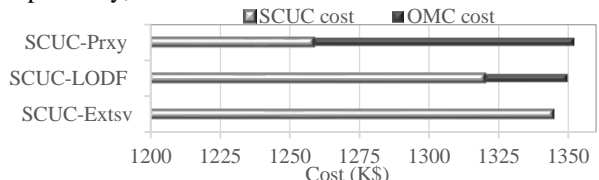

Fig. 2: Final costs comparison for $N-1$ reliable solutions.

Fig. 2 compares the final costs for the different market auction models. This cost includes the SCUC cost and OMC cost. OMC is not performed for SCUC-Extsv as this model explicitly represents contingencies using recourse decision variables and produces an $N$-1 reliable solution. The solution of the SCUC-Extsv market auction model has the lowest final cost (benchmark solution) since its scheduled reserve is deliverable in post-contingency states. The SCUC-LODF results in higher SCUC cost compared to the SCUC-Prxy, but it requires less discretionary changes or uneconomic adjustments (OMC actions) to achieve $N$-1 reliability; thus, the SCUC-LODF results in less OMC cost. From the reliability point of view, it can be concluded that the SCUC-LODF provides a solution that is closer to $N$-1 reliable SCUC-Extsv solution.

LMPs of the market models are studied in Fig. 3 for hour 22 across the buses. Based on this figure, as the market model moves away from SCUC-Prxy toward capturing more accurate representation of the contingency events, the prices are increased from bus #67 to bus #109, and also are mostly higher in bus #1-67. The difference between prices is due to the new elements of LMP, i.e., marginal security elements, which represent the value of reserve provision in the modeled contingencies. More specifically, the deterministic model, which utilizes proxy reserve to achieve $N$-1 reliability, does not capture the true value of achieving $N$-1 reliability due to the fact that the obtained market LMPs do not adequately reflect the value of delivering reserve in the post-contingency state. Accordingly, the LMPs tend to be lower in this model. In this case, the new committed units after OMC may not be fully compensated for providing ancillary services, which can be a reason of *missing money* issue. This will cause a natural unfairness in market strategy as market participants might not be compensated fairly with this mechanism. On the other hand, the SCUC-Extsv inherently captures the different values of reserves offered by various entities, as it reflects the value of delivering reserve in the post-contingency state on a locational basis, so the SLMPs tend to be higher. This result occurs because the model explicitly checks to see whether the reserve is deliverable for each contingency. *Overall, these analyses confirm that with more accurate representation of contingencies in the market auction models, the reliability and associated products are priced more accurately.* This would result in fair and accurate market signals for market participants and improve overall market efficiency.

Fig. 4 compares the market auction models with respect to market settlements. It can be seen that the generators revenue and load payment have the highest values with the SCUC-Extsv model, while they have the lowest value with the SCUC-Prxy. The generation rent is calculated from subtracting variable cost of units from their revenues, which also increases as the models have more explicit and accurate representation of the contingency events. From these results, it can be said that more accurate modeling of $N$-1 requirement in market models results in increased profit of generators.

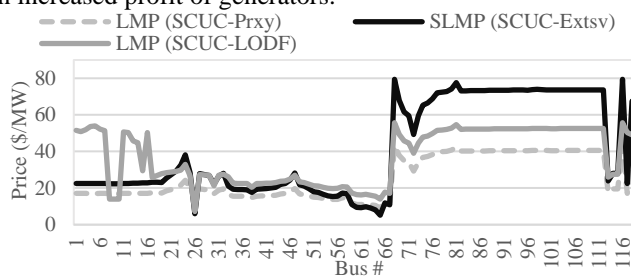

Fig. 3: Pricing comparison of SCUC-Prxy, SCUC-LODF, and SCUC-Extsv.

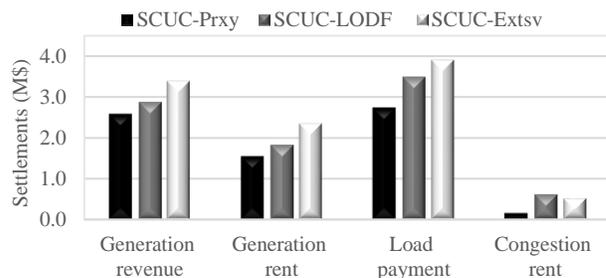

Fig. 4: Settlements for different market action models.

## IV. OBJECTIVE FUNCTION DESIGN AND FORMULATION EVALUATION FOR STOCHASTIC MARKET MODELS

ESCUC model presented by (24)-(35) minimizes the expected cost of all scenarios (called ESCUC-*expected* model in rest of this paper), as presented in (24). However, during emergency conditions in RT, the system operators implement corrective actions, which are aimed to eliminate violations as quickly as possible (not to minimize operation cost) in order to recover from the contingency and to regain $N$-1 reliability. These corrective actions may not necessarily be the lowest cost options. Thus, minimizing post-contingency cost in the DA may result in a solution, which anyways will not be fully implemented. This discrepancy can result in inefficiency of stochastic market model with expected cost minimization objective. Furthermore, minimizing expected cost may result in deviation of the base-case schedule, which has the highest probability of occurrence, from its optimal solution. In short, such models may generate operational schedules with higher base-case cost with no guarantee of reducing operating costs during contingencies. Additionally, inaccurate estimation of the probability of the asset outages can lead to different pricing and market outcomes.

The aforementioned issues create a need for detailed examination of the objective function design for the stochastic market models. An alternative option for ESCUC is to minimize only the base-case costs including generator production costs, the startup costs, and the shutdown costs as shown in (37), with the same set of constraints as (25)-(35).

$$minimize \sum_g \sum_t \left(c_g^p P_{g0t} + c_g^{NL} u_{gt} + c_g^{SU} v_{gt} + c_g^{SD} w_{gt}\right) \quad (37)$$

The above model for stochastic market is called ESCUC-*base* market model throughout this paper. This model minimized (37) while searching for a feasible solution for pre- and post-contingency states. This paper proposes a framework to identify an effective stochastic market design by comparing two ESCUC models, i.e., ESCUC-*expected* and ESCUC-*base*, from the aforementioned aspects. The detailed analyses are presented in the following sub-sections.

*A. Realized N-1 final operating cost*

As discussed, intention during an emergency condition is not to minimize cost; rather, the goal is to recover from the event as quickly as possible by minimizing violation. In this section, contingency analysis with violation minimization is performed, which mimics the operator's actions in emergency conditions. The realized *N*-1 operating costs for the two ESCUC models are calculated using the dispatch from contingency analysis tool; these costs reflect the corrective generation dispatch actions after the *N*-1 contingencies. Then, the realized *N*-1 costs for the dispatch from two ESCUC models are compared. Fig. 5 demonstrates the procedure performed for this analysis.

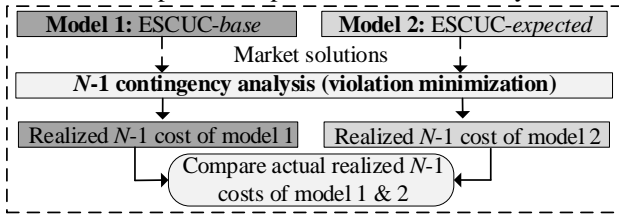

Fig. 5: Procedure of comparison of two models for actual realized *N*-1 costs.

The *N*-1 contingency analysis tool is a linear programming problem which is solved independently at each time period $t$ for each operating state $c \in C_g, C_k$. The formulation for contingency analysis is given below.

$$minimize \sum_n (LS_n^+ + LS_n^-) \quad (39)$$
$$-P_g \leq (\bar{r}_g - \bar{P}_g)\bar{u}_g N1_g, \forall g \quad (40)$$
$$P_g \leq (\bar{r}_g + \bar{P}_g)\bar{u}_g N1_g, \forall g \quad (41)$$
$$P_g^{min}\bar{u}_g N1_g \leq P_g \leq P_g^{max}\bar{u}_g N1_g, \forall g \quad (42)$$
$$P_n^{inj} = \sum_{g \in g(n)} P_g - Load_n + LS_n^+ - LS_n^-, \forall n \quad (43)$$
$$\sum_n P_n^{inj} = 0 \quad (44)$$
$$-P_k^{max,c} N1_k \leq \sum_n PTDF_{nk}^{ref} P_n^{inj} \leq P_k^{max,c} N1_k, \forall k \quad (45)$$
$$LS_n^-, LS_n^+ \geq 0, \forall n \quad (46)$$

Positive slack variables, i.e., $LS_n^-$ for load shedding and $LS_n^+$ for load surplus, indicate the post-contingency security violations. Consequently, the contingency analysis objective (39) is to minimize the load shed and the load surplus, when an outage occurs. Constraints (40) and (41) restrict the deviation of the power generation from the pre-contingency to the post-contingency by the scheduled reserve obtained from the DA ESCUC models. The generator output limit constraint in post-contingency state is represented by (42). The node balance constraint in the post-contingency state is ensured by (43), while (44) ensures power balance at system level. Constraint (45) limits the post-contingency transmission line flows to be within the emergency limits for generation and transmission contingencies. In model (39)-(46), only one $N1_k$ and $N1_g$ is set to zero while the rest of $N1_k$ and $N1_g$ are equal to one.

*B. Impacts of imprecise estimation in probabilities*

The second issue that should be investigated when it comes to design of a stochastic market model is the implications of inaccuracy in the estimation of probability of outages. These analyses are also very necessary to be performed as it is difficult to exactly estimate the probability of outages, which itself can lead to different pricing implications and market solutions. In order to realize the impact of this inaccuracy, procedure shown in Fig. 6 is proposed in this paper.

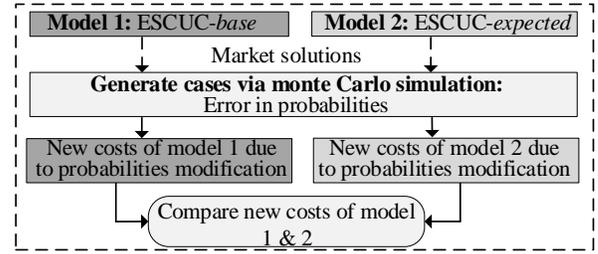

Fig. 6: Procedure of analysis of imprecise probabilities estimation.

*C. Testing & Results of objective function design*

IEEE 118-bus test system that was explained in detail at section III.A is used to perform the simulations. First, *ESCUC-base* and *ESCUC-expected* models are solved with relative MIP gap set to 0% to compare their benchmark solutions. The formulations are evaluated based on DA operational scheduling cost and the realized operation cost during *N*-1 contingency scenarios (i.e., contingency analysis with violation minimization). Fig. 7 (a) compares the two different costs, i.e., original costs versus realized *N*-1 costs for two market auction models. It is clear that the original DA expected costs of the ESCUC-*expected* model are lower than those of ESCUC-*base* model, as the objective function of the ESCUC-*expected* is to minimize the costs over all scenarios while the ESCUC-*base* minimizes just the base-case costs. However, the realized *N*-1 costs of two models are almost the same (the difference is only 0.002 percent). These results reveal that minimizing post-contingency costs in the ESCUC-*expected* does not represent operators' actions and may not result in lower *N*-1 realized costs in the emergency conditions. Similar results are obtained when the scenario costs of the two models are compared as shown in Fig. 7 (b). It is pertinent to note that the scenario costs include the expected variable cost of generators for post-contingency scenarios (all scenarios excluding the base-case scenario).

Moreover, the industry practice of considering a non-zero MIP gap is implemented here to achieve 30 various solutions (all within 1% MIP gap) for the two ESCUC models, for the sake of further comparison. By pairing the solutions of two ESCUC models, the total number of pairs is equal to

30×30=900, each of which includes a possible markets outcome from two different models to be compared.

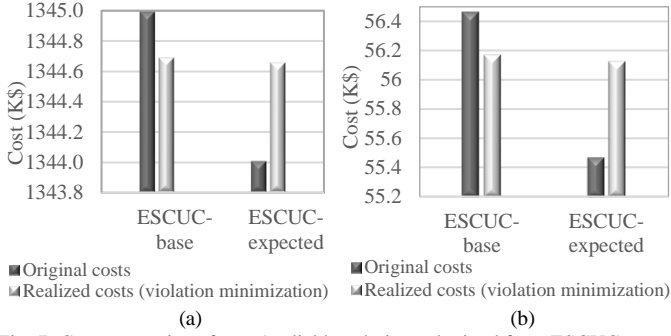

Fig. 7: Cost comparison for *N*-1 reliable solutions obtained from ESCUC-*expected* and ESCUC-*base* models: (a) expected costs (b) scenario costs.

Table I lists the percentage of pairs that the ESCUC-*base* model results in lower cost compared to ESCUC-*expected*. It can be observed that, as expected, the DA base-case cost of ESCUC-*base* is lower in 85 percent of pairs compared to the ESCUC-*expected* model. Fig. 8 (a) presents the histogram of costs difference calculated from subtracting the base-cost cost of ESCUC-*base* from that of ESCUC-*expected*. It can be seen that the density of the pairs cost difference tends to be toward positive values. Moreover, Table I presents that the original DA expected costs of ESCUC-*base* are lower in 30 percent of the pairs as expected (see Fig. 8 (b) for histogram illustration of difference in original DA expected costs). Finally, Table I shows that in almost 86 percent of pairs, the realized *N*-1 costs of ESCUC-*base* model are less than those of ESCUC-*expected* model during *N*-1 contingency scenarios. Fig. 8 (c) illustrates realized *N*-1 costs difference (i.e., realized $\{N-1\}\ \text{costs}_{\text{ESCUC}-expected}$ − realized $\{N-1\}\ \text{costs}_{\text{ESCUC}-base}$) for the 900 pairs.

TABLE I: PERCENTAGE OF PAIRS WITH LOWER COST FOR ESCUC-*BASE* MODEL COMPARED TO ESCUC-*EXPECTED* MODEL.

| Type of cost | Percentage |
| --- | --- |
| DA base-case costs | 85% |
| Original DA expected costs | 30% |
| Realized *N*-1 costs (violation minimization) | 86% |

The summarized results in Table I as well as Fig. 8 confirm that the realized *N*-1 costs from the DA dispatch solution of ESCUC-*base* are lower than the realized *N*-1 costs of ESCUC-*expected* in most of the pairs regardless of *N*-1 cost minimization in the ESCUC-*expected*. Moreover, the base-case costs of ESCUC-*base* are lower than those of the ESCUC-*expected* in the most of pairs. As one can see, the ESCUC-*base* performs better in general compared to the ESCUC-*expected* model based on the base-case costs and realized *N*-1 final costs.

Moreover, the impact of inaccuracy in the contingency probability estimation is evaluated. It is assumed that there is an error in the estimation of probabilities that follows a Gaussian distribution with zero mean, and the standard deviation of 20% and 40%. For each standard deviation, 2000 cases are generated, each of which includes a set of 231 contingency probabilities (for generator and non-radial transmission line contingencies). Then, the probability of base case is calculated through following equation.

$$\pi_{BC,s} = 1 - \sum_{c\in C_k} \pi_{c,s} - \sum_{c\in C_G} \pi_{c,s}, \forall s \qquad (47)$$

where $s$ is the index of cases, and $\pi_{BC,s}$ and $\pi_{c,s}$ are probability of base-case scenario and contingency event $c$ in case $s$. Since the range of probability of base-case scenario is more known for the system operators for a specific electric system, the cases that lead to a base-case probability out of the range 0.948 and 0.944 have been eliminated (perfect estimation has a base-case probability of 0.946). The 2000 cases are applied based on the procedure presented in Fig. 6 to each of the 30 solutions of ESCUC-*base* and ESCUC-*expected* models mentioned earlier. Therefore, there is 30×2000=60,000 solutions (costs) for each market model. By pairing the solutions of two different stochastic market model, total number of pairs is equal to 3,600,000,000, i.e., 60,000×60,000. It is worth mentioning that each pair has two market outcomes, one from ESCUC-*base* model and one from ESCUC-*expected* that can be compared.

TABLE II: EFFECTS OF ERROR IN ESTIMATION OF PROBABILITIES ON PERCENTAGE OF PAIRS THAT ESCUC-BASE MODEL HAS LOWER COST COMPARED TO ESCUC-*EXPECTED* MODEL.

| | Perfect estimation | 20% error in estimation | 40% error in estimation |
| --- | --- | --- | --- |
| % of pairs with lower original expected cost | 30 | 36.6 | 43.0 |
| % of pairs with lower scenario cost | 0 | 7.8 | 22.2 |

Table II presents the impacts of error in estimation of probabilities on the percentage of the pairs that ESCUC-*base* model has lower cost (original expected cost and scenario cost) compared to the ESCUC-*expected* model. From Table II, it can be seen that the percentage of pairs where ESCUC-*base* model has lower original expected cost increase from 30% to 36.6% and 43% as the accuracy in estimation of probabilities moves from being perfect to have 20% and 40% estimation errors, respectively. The percentage of pairs in which the ESCUC-*base* model has lower scenario cost in comparison with the other model increases from 0% for perfect estimation to 7.8% and 22.2% for 20% and 40% error in estimation, respectively. These results demonstrate that after the solutions of two models are affected by the inaccuracy in probabilities estimation, the likelihood that ESCUC-*base* outperforms the other model in having less original expected cost and scenario cost increases.

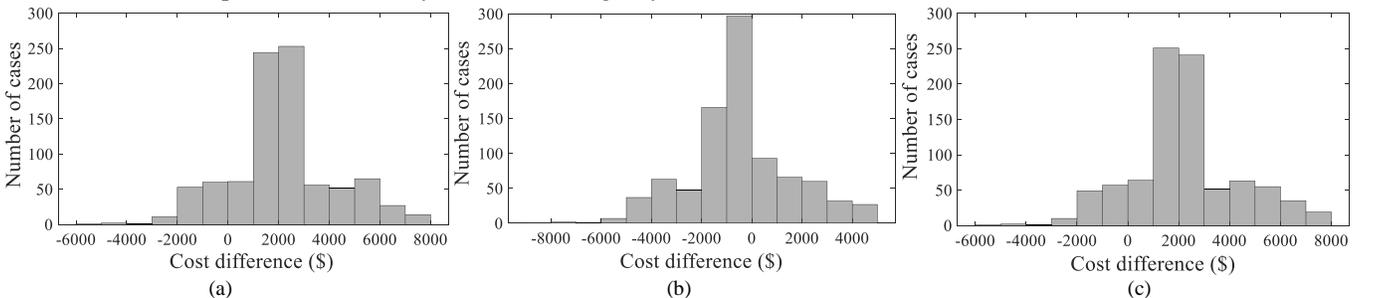

Fig. 8: Histogram of cost difference ($cost_{\text{ESCUC}-expected} - cost_{\text{ESCUC}-base}$) of the pairs: (a) base-case costs (b) original expected costs (c) realized *N*-1 costs.

V. CONCLUSION

A comprehensive framework incorporating various procedures was proposed in this paper to: (i) conduct a fair comparison of pricing and settlements between different market models that ensure different levels of security, and (ii) examine efficient objective formulation for stochastic market design.
To compare various market models: (i) the concept of securitized LMP was developed for ESCUC model, (ii) an OMC procedure was implemented on the solution of market models with proxy reserve requirement and with LODF to obtain *N*-1 reliable dispatch. The ISO practice of calculating market settlements based on the original market prices and *N*-1 reliable schedule after performing OMC was implemented. Newly committed generators during OMC do not have direct impact on the value of LMP for their location. At this stage, it is unclear whether this mechanism (a SCUC model following with OMC) enables opportunities for market exploitation. Using this mechanism, the market model is not purely a pool; instead, it is a combination of a pool and pay-as-bid model. This practice is limited and not transparent for all market participants; therefore, the market participants will not change their bidding strategy accordingly (as they would in a pay-as-bid model). Although this is an accepted market manipulation, some participants might receive less than deserved benefits and some might receive more. However, with more accurate representation of contingencies in the ESCUC compared to SCUC models with approximation on *N*-1 security criteria, *N*-1 grid security requirements are originally captured, thereby, the value of service (contingency-based reserve) provided by generators is reflected in the LMPs to achieve grid security. In other words, if the market SCUC includes the reliability criteria more adequately, prices can better reflect the true marginal cost associated with the provision of the reliable electricity.

Furthermore, it was shown that the stochastic market design with expected objective function does not give solutions that ensure minimum realized operating costs at *N*-1 contingency states. Instead, the stochastic market design with base-case objective function had better performance compared to the market model with expected objective function in terms of the base-case costs and realized *N*-1 costs. Moreover, inaccuracy in estimated probability results in larger differences in the original DA expected costs and the costs of scenarios of ESCUC-*base* and ESCUC-*expected*, where ESCUC-*base* further outperforms ESCUC-*expected*. It can be concluded that evidently the stochastic market design with base-case objective function can be more efficient compared to the stochastic market design with expected objective function.

REFERENCES

[1] National Academy of Engineering, "Modernizing and Protecting the Electricity Grid," Spring 2010. [Online]. Available: https://nae.edu/18627/Modernizing-and-Protecting-the-Electricity-Grid.
[2] NERC, "Reliability concepts," Mar. 2016. [Online]. Available: http://www.nerc.com/files/concepts_v1.0.2.pdf.
[3] M. Sahraei-Ardakani and K. W. Hedman, "Day-Ahead Corrective Adjustment of FACTS Reactance: A Linear Programming Approach," *IEEE Trans. Power Syst.*, vol. 31, no. 4, pp. 2867–2875, Jul. 2016.
[4] Y. Chen, P. Gribik, and J. Gardner, "Incorporating Post Zonal Reserve Deployment Transmission Constraints Into Energy and Ancillary Service Co-Optimization," *IEEE Trans. Power Syst.*, vol. 29, no. 2, pp. 537–549, Mar. 2014.
[5] M. Abdi-Khorsand, M. Sahraei-Ardakani, and Y. Al-Abdullah, "Corrective transmission switching for N-1-1 contingency analysis," *IEEE Trans. Power Syst.*, vol. 32, no. 2, pp. 1606–1615, Mar. 2017.
[6] CAISO, "Market Performance Metric Catalog," Nov. 2019. [Online]. Available: http://www.caiso.com/Documents/MarketPerformanceMetricCatalogforNovember2019.pdf.
[7] B. F. Hobbs, M.-C. Hu, J. G. Inon, S. E. Stoft, and M. P. Bhavaraju, "A Dynamic Analysis of a Demand Curve-Based Capacity Market Proposal: The PJM Reliability Pricing Model," *IEEE Trans. Power Syst.*, vol. 22, no. 1, pp. 3–14, Feb. 2007.
[8] P. Gribik, W. Hogan, and S. Pope, "Market-Clearing Electricity Prices and Energy Uplift," 2007. [Online]. Available: http://www.lmpmarketdesign.com/papers/Gribik_Hogan_Pope_Price_Uplift_123107.pdf.
[9] J. M. Arroyo and F. D. Galiana, "Energy and reserve pricing in security and network-constrained electricity markets," *IEEE Trans. Power Syst.*, vol. 20, no. 2, pp. 634–643, May 2005.
[10] F. Bouffard, F. D. Galiana, and A. J. Conejo, "Market-clearing with stochastic security-part I: formulation," *IEEE Trans. Power Syst.*, vol. 20, no. 4, pp. 1818–1826, Nov. 2005.
[11] F. Bouffard, F. D. Galiana, and A. J. Conejo, "Market-clearing with stochastic security-part II: case studies," *IEEE Trans. Power Syst.*, vol. 20, no. 4, pp. 1827–1835, Nov. 2005.
[12] S. Wong and J. D. Fuller, "Pricing Energy and Reserves Using Stochastic Optimization in an Alternative Electricity Market," *IEEE Trans. Power Syst.*, vol. 22, no. 2, pp. 631–638, May 2007.
[13] B. Stott, J. Jardim, and O. Alsac, "DC Power Flow Revisited," *IEEE Trans. Power Syst.*, vol. 24, no. 3, pp. 1290–1300, Aug. 2009.
[14] CAISO, "Draft final proposal: Generator contingency and remedial action scheme modeling," Jul. 2017. [Online]. Available: https://www.caiso.com/Documents/DraftFinalProposal-GeneratorContingencyandRemedialActionSchemeModeling_updatedjul252017.pdf.
[15] N. G. Singhal, N. Li, and K. W. Hedman, "A Data-Driven Reserve Response Set Policy for Power Systems With Stochastic Resources," *IEEE Trans. Sustain. Energy*, vol. 10, no. 2, pp. 693–705, Apr. 2019.
[16] C. Li, K. W. Hedman, and M. Zhang, "Market pricing with single-generator-failure security constraints," *IET Gen. Trans. & Dist.*, vol. 11, no. 7, pp. 1777–1785, 2017.
[17] R. Fernández-Blanco, Y. Dvorkin, and M. A. Ortega-Vazquez, "Probabilistic Security-Constrained Unit Commitment With Generation and Transmission Contingencies," *IEEE Trans. Power Syst.*, vol. 32, no. 1, pp. 228–239, Jan. 2017.
[18] C. J. López-Salgado, O. Añó, and D. M. Ojeda-Esteybar, "Stochastic Unit Commitment and Optimal Allocation of Reserves: A Hybrid Decomposition Approach," *IEEE Trans. Power Syst.*, vol. 33, no. 5, pp. 5542–5552, Sep. 2018.
[19] V. Guerrero-Mestre, Y. Dvorkin, R. Fernández-Blanco, M. A. Ortega-Vazquez, and J. Contreras, "Incorporating energy storage into probabilistic security-constrained unit commitment," *IET Gen. Trans. & Dist.*, vol. 12, no. 18, pp. 4206–4215, 2018.
[20] J. Wang, M. Shahidehpour, and Z. Li, "Contingency-Constrained Reserve Requirements in Joint Energy and Ancillary Services Auction," *IEEE Trans. Power Syst.*, vol. 24, no. 3, pp. 1457–1468, Aug. 2009.
[21] Z. Guo, R. L. Chen, N. Fan, and J. Watson, "Contingency-Constrained Unit Commitment With Intervening Time for System Adjustments," *IEEE Trans. Power Syst.*, vol. 32, no. 4, pp. 3049–3059, Jul. 2017.
[22] Y. M. Al-Abdullah, M. Abdi-Khorsand, and K. W. Hedman, "The Role of Out-of-Market Corrections in Day-Ahead Scheduling," *IEEE Trans. Power Syst.*, vol. 30, no. 4, pp. 1937–1946, Jul. 2015.
[23] University of Washington, "Power systems test case archive," 1999. [Online]. Available: http://www.ee.washington.edu/research/pstca/index.html.
[24] A. T. Saric, F. H. Murphy, A. L. Soyster, and A. M. Stankovic, "Two-Stage Stochastic Programming Model for Market Clearing With Contingencies," *IEEE Trans. Power Syst.*, vol. 24, no. 3, pp. 1266–1278, Aug. 2009.